\documentstyle[epsfig]{mn2e}

\title[The signature of Pop III stars]{The chemical signature of surviving Population III stars in the Milky Way}

\author[J. L. Johnson]{Jarrett L. Johnson\thanks{Email:
    jlj@lanl.gov} \\
X Theoretical Design, Los Alamos National Laboratory, Los Alamos, NM  87545, USA}

\pubyear{2015}

\begin{document}
\maketitle

\topmargin-1cm

\begin{abstract}
Cosmological simulations of Population (Pop) III star formation
suggest that the primordial initial mass function may have extended 
to sub-solar masses.  If Pop III stars with masses $\le$ 0.8
M$_{\odot}$ did form, then they should still be present in the Galaxy
today as either main sequence or red giant stars.  Despite broad
searches, however, no primordial stars have yet been identified.  It
has long been recognized that the initial metal-free nature of
primordial stars could be masked due to accretion of metal-enriched
material from the interstellar medium (ISM).  Here we point out that
while gas accretion from the ISM may readily occur, the accretion of
dust from the ISM can be prevented due to the pressure of the
radiation emitted from low-mass stars.  This implies a possible unique
chemical signature for stars polluted only via accretion, namely an
enhancement in gas phase elements relative to those in the dust phase.
Using Pop III stellar models, we outline the conditions in which this
signature could be exhibited, and we derive the expected signature for
the case of accretion from the local ISM.  Intriguingly, due to the
large fraction of iron depleted into dust relative to that of carbon
and other elements, this signature is similar to that observed in many
of the so-called carbon-enhanced metal-poor (CEMP) stars.  We
therefore suggest that some fraction of the observed CEMP stars may,
in fact, be accretion-polluted Pop~III stars. 
\end{abstract}


\section{Introduction}
For decades it has been a critical open question what were the first
stars, and what was there fate (e.g. Bond 1981).  Increasingly broad
and sensitive surveys have been carried out in the halo (e.g. Cayrel
et al. 2004; Caffau et al. 2011; Keller et al. 2014) and bulge (e.g. Schlaufman \& Casey 2014) of
the Galaxy, as well as in nearby dwarf galaxies (e.g. Kirby et
al. 2011; Frebel et al. 2014) in search of the most primitive stars (see e.g. Beers \&
Christlieb 2005; Frebel 2010 for reviews).  While
these surveys have uncovered a trove of extremely metal-poor stars 
that provide invaluable clues to the nature of the first stars, as of
yet there have been found no stars with a truly primordial composition (i.e. no metals).  
 
This null result in the hunt for Population (Pop) III stars is
increasingly in tension with state-of-the-art cosmological simulations
suggesting that low-mass Pop III stars may have formed in the early universe
(e.g. Clark et al. 2011; Greif et al. 2011; Dopcke et al. 2013; Bromm 2013; Susa et al. 2014; Stacy \& Bromm
2014; Greif 2014; Hirano et al. 2014; Machida \& Nakamura 2015) and that they
may still reside in the Galaxy today (e.g. Madau et
al. 2008; Gao et al. 2010; Tumlinson et al. 2010; Karlsson et al. 2013; Hartwig et al. 2014).  One clear resolution to this tension
between theory and observation emerges if low-mass Pop III stars 
accrete metals from the interstellar medium (ISM) during their long
lives traversing the Galaxy (Yoshii 1981; Iben 1983; Frebel et al. 2009; Komiya et
al. 2010, 2015; Johnson \& Khochfar 2011).  Such accretion events would
result in the pollution of the stellar surface with heavy elements and 
mask the primordial nature of a Pop III star.  
 
Here we consider the expected chemical signature produced by such accretion
events.  In particular, we show that a clear signature could be imprinted due to the 
fact that, owing to the force of the radiation emitted from low-mass
stars, dust is often not accreted onto such stars even while gas is
readily accreted.  In Section 2, we estimate the impact of stellar
radiation on the dynamics of dust grains.  In Sections 3 and 4, we describe
the conditions under which dust and gas are segregated in accretion
flows due to the influence of stellar radiation.  In Section 5, we estimate the
expected chemical signatures of first and second generation stars that
are enriched via accretion of gas and dust from an ISM with heavy element
abundance ratios and dust depletion properties similar to those of the local ISM.  
Finally, we give our conclusions and offer a brief discussion in Section 6.

  \begin{figure}
   \includegraphics[angle=270,width=3.4in]{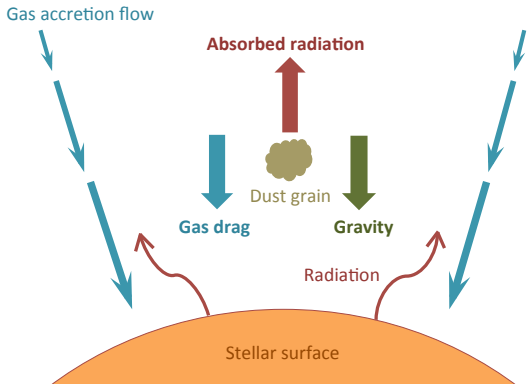}
   \caption{The forces acting on a dust grain in an accretion flow
     onto a low-mass star:  the inward gravitational pull of
     the star ({\it green}), the outward force due to absorption of
     stellar radiation ({\it red}), and the inward drag on the grain
     due to collisions and Coulomb interactions (see
     Section 3) with the accreting gas particles ({\it blue}).
     If a sufficient amount of radiation is absorbed by the grain, it
     will be repelled from the star, even as gas particles are readily accreted.}
   \end{figure}

\section{The Eddington limit for dust}
Here we consider the forces at play during the accretion of
dust-enriched material from the ISM onto a star, as illustrated in
Figure 1.  In particular, we shall concern ourselves with accretion
onto low-mass ($\le$ 0.8 M$_{\odot}$) Pop~III stars and low-mass, low-metallicity 
Pop~II protostars.  While the radiation emitted from these objects
couples weakly to gas, it can be readily absorbed by dust grains.  In
turn, this implies that the dynamics of dust grains can be quite
different from that of the gas, as has been discussed previously in the context
of various types of metal-enriched stars (e.g. Voshchinnikov \& Il'in
1983; Ivezic \& Elitzur 1995; 2010;
Netzer 2007) and in pre-stellar cores (Whitworth \& Bate 2002) and
diffuse clouds (Weingartner \& Draine 2001b) exposed to the radiation
field in the diffuse ISM.

We shall define the Eddington ratio for dust as the ratio of the
outward radiative force on a grain to the inward gravitational force
on the grain.  As shown in Figure 1, if a sufficient amount of
radiation is absorbed by a grain, the Eddington ratio can exceed
unity, in which case the grain may be repelled from the star even as
gas is accreted onto it.  This is, however, provided that the inward drag
force due to collisions and Coulomb interactions with gas particles is sufficiently small, as we
discuss in the next Section.

The Eddington ratio for a given dust grain depends on its size,
density and the efficiency with which it absorbs radiation.  
 Here we consider a range of dust grain densities
between 1 and 3 g cm$^{-3}$, roughly bracketing the densities of 
porous and compact interstellar dust grains, respectively
(e.g. Ossenkopf 1993; Mathis 1996;
Iat{\' i} et al. 2001; Weingartner \&
Draine 2001a;  Dullemond \& Dominik 2005; Voshcninnikov et al. 2005;
Shen et al. 2008; Heng \& Draine 2009), and we
adopt the wavelength-dependent absorption efficiencies for both
graphite and silicate grains presented in Draine \& Lee
(1984). 

 We note that the absorption efficiencies of porous
  grains may be somewhat different than those of the compact grains
 that we adopt here  (e.g. Mukai et al. 1992; Tazaki \& Nomura 2015),
 although it has been shown that porous (aggregate) grains composed of
 silicates and graphite have absorption efficiencies that differ by
less than a factor of $\sim$ 2 over the range of porosities that we
consider for interstellar grains (for effective radii of $\sim$ 0.1
$\mu$m; see Shen et al. 2008).  For
simplicity and given the other large uncertainties in our calculations
 (e.g. ISM properties, dust properties and depletion factors over
 the history of the Galaxy) we adopt the same absorption efficiencies 
for porous and compact grains. 

These radiation absorption efficiencies are dependent on the
wavelength of the radiation and so on the spectrum of the radiation
emitted by the star.  For the stellar radiation we adopt simple black body spectra.
We use effective temperatures and luminosities appropriate for a 
0.8 M$_{\odot}$ Pop III star, as presented in Siess et al. (2002).  Specifically, we choose 
a temperature of 6500 K and a luminosity of 5 L$_{\odot}$ for the main
sequence (MS) stage, and a temperature of 5500 K and a luminosity of 50
L$_{\odot}$ for the red giant (RG) stage.\footnote{Other low-mass Pop III stellar models give similar
  values (Marigo et al. 2001; Picardi et al. 2004; Suda et al. 2007).}   
For Pop~II protostellar radiation, we adopt values for the temperature and luminosity in the range of values derived from 
observations as presented in Dunham et al. (2014).  We choose a
temperature of 1000 K, near the upper end of the distribution of
protostellar temperatures observed in the Galaxy, as we expect
sub-solar metallicity protostars to be hotter, on average, than solar
metallicity protostars in the Galaxy today, due to the lower opacity
of metal-poor stellar envelopes (see e.g. Marigo et al. 2001; Schaerer
2002; Suda et al. 2007).  Consistent 
with this temperature, we choose a luminosity of
100 L$_{\odot}$.\footnote{In principle, at these luminosities it is possible for dust grains
  to be sublimated in the accretion flows that we consider, although
  in Appendix A we show that this is may be unlikely to occur.}

The Eddington ratios for dust grains, as functions of their radii
(referred to as size in the Figures),
are shown in Figures 2, 3 and 4, for Pop III MS stars, Pop III RGs and
Pop II protostars, respectively.  In each Figure, we present the
Eddington ratios for compact and porous grains, for both graphite and
silicates.  In each Figure, it is clear that less dense (more porous)
grains have higher Eddington ratios.  This follows from the fact that, for a given size, 
denser grains are more massive and so experience a stronger
gravitational pull toward the star.    

Comparing the Figures, it is also clear that the Eddington ratios are
much higher for the more luminous Pop~III RG and Pop~II protostellar
cases, than for the less luminous Pop~III MS case.  We note, in
particular, that the Eddington ratio is $\sim$ 10 times higher in the
Pop~III RG case than in the Pop~III MS case; this is expected, given
that the luminosities we have adopted for these cases differ by just
this factor and that the effective temperatures are similar in both
cases.

Finally, we note that the Eddington ratios are higher for graphite
than for silicates.  This follows directly from the fact that the
absorption efficiencies for graphite grains are significantly higher than those for
silicates, in the wavelength range in which the stars are most
strongly emitting.  For the black body spectra we adopt, the emission
peaks at 0.44, 0.52 and 2.89 $\mu$m for the Pop~III MS (6500 K),
Pop~III RG (5500 K) and Pop~II protostellar (1000 K) cases,
respectively.  It is at these and larger wavelengths that the
absorption efficiency of graphite is especially higher than that of
silicates (see Figures 4 and 5 of Draine \& Lee 1984).

A large fraction of the dust grains shown in the Figures have
Eddington ratios exceeding unity, implying that they may not be
accreted onto the low-mass stars and protostars that we
consider.\footnote{For extremely high Eddington ratios, in particular
  those exceeding the gas to dust mass ratio (which may be, e.g., $\sim$
  100), it is possible that the radiatively-accelerated dust grains
  may sweep up the gas as they are expelled from the star.  In such
  cases, neither gas nor dust would be accreted.  As we find, in
  general, much lower Eddington ratios, our calculations suggest this
  is uncommon.}
However, as shown in Figure 1, the force of gas drag must also be overcome
in order for grains to avoid accretion; we consider this
effect in the next Section.

  \begin{figure}
   \includegraphics[angle=0,width=3.4in]{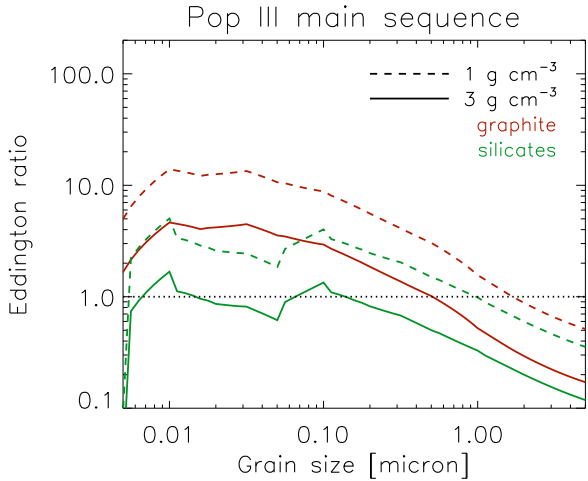}
   \caption{The Eddington ratio for dust grains in an accretion flow
     onto a 0.8 M$_{\odot}$ MS Pop III star with a surface temperature
     of 6500 K and a luminosity of 5 L$_{\odot}$, shown as a function
     of grain size.  The dotted lines correspond to relatively porous
     grains with a density of 1 g cm$^{-3}$, while the solid lines
     correspond to more compact grains with a density of 3 g
     cm$^{-3}$.  An Eddington ratio greater than unity implies that
     the outward radiative force exceeds the inward gravitational
     force, and that the grains can be repelled from the star instead
     of being accreted along with the gas.  For a given grain size, more porous grains experience a weaker gravitational pull toward the star, leading to higher
     Eddington ratios.}
   \end{figure}

  \begin{figure}
   \includegraphics[angle=0,width=3.4in]{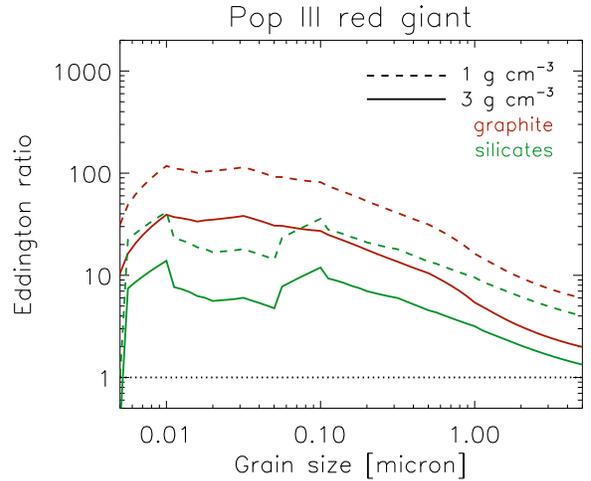}
   \caption{The same as Figure 2 but for the case of accretion onto a
     0.8 M$_{\odot}$ Pop III RG star with a surface temperature of
     5500 K and a luminosity of 50 L$_{\odot}$.  The larger luminosity 
     of the RG, as compared to the MS star
     shown in Figure 2, implies larger radiative forces on the grains,
     and thus larger Eddington ratios.}
   \end{figure}

  \begin{figure}
   \includegraphics[angle=0,width=3.4in]{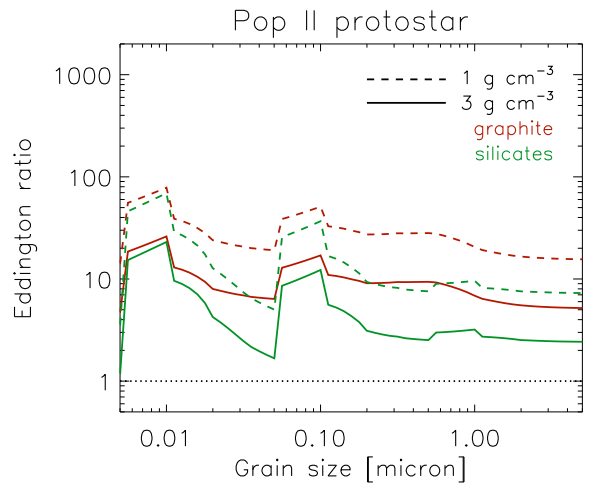}
   \caption{The same as Figures 2 and 3 but for the case of accretion
     onto a 0.8 M$_{\odot}$ Pop~II protostar with a temperature of 1000 K and a
     luminosity of 100 L$_{\odot}$.}
   \end{figure}

\section{The effect of gas drag}
Here we consider the effects of particle collisions and Coulomb drag on the dynamics of dust grains
in accretion flows.  We include these forces in the overall force
balance shown in Figure 1, and determine the conditions under which
dust grains will not be accreted along with the gas.  

To estimate the force due to collisional gas drag $F_{\rm collision}$
and Coulomb drag $F_{\rm Coulomb}$ we adopt the equation below
from Draine \& Salpeter (1979)\footnote{We note that other authors have also described similarly
the effect of gas drag on radiation-driven gas and dust segregation
(e.g. Baines et al. 1965; Simpson et al. 1980).} :

\begin{eqnarray}
F_{\rm drag} & = & F_{\rm collision} + F_{\rm Coulomb}   \nonumber \\
& = & 2 \pi r^2_{\rm grain} k_{\rm B} T      \nonumber \\
& \times & n_{\rm H} \left[ G_{\rm 0}(s) + f_{\rm e} \psi^2
  {\rm ln}(\Lambda) G_{\rm 2}(s) \right]  \nonumber \\
& = & 2 \pi r^2_{\rm grain} k_{\rm B} T  n_{\rm H} G_{\rm 3} \mbox{\ ,}
\end{eqnarray}
where $T$ is the gas temperature, $k_{\rm B}$ is Boltzmann's constant,
$r_{\rm grain}$ is the radius of dust grains, and $n_{\rm H}$ is the
number density of hydrogen nuclei.  For our calculations (see Figure
5) we assume a hydrogen gas with a free electron (or ionized
fraction) fractions of $f_{\rm e}$ = 10$^{-2}$ and 10$^{-4}$, bracketing values that are appropriate for the
cold, dense ISM (e.g. Weingartner \& Draine 2001c; Draine 2003).   As
reflected in equation (1), for our calculations we assume that grains are spherical, although
there are significant uncertainties related to departures from
spherical symmetry.   We have also adopted here, again following 
Draine \& Salpeter (1979), the following formulae:

\begin{equation}
\psi= \frac{e U}{k_{\rm B} T} \mbox{\ ,}
\end{equation}

\begin{equation}
\Lambda = \frac{3}{2 r_{\rm grain} e \psi} \left(k_{\rm B} T / \pi
  n_{\rm e} \right)^{\frac{1}{2}} \mbox{\ ,}
\end{equation}

\begin{equation}
s = \left(m_{\rm grain} c^2_{\rm s} / 2 k_{\rm B} T
\right)^{\frac{1}{2}} \mbox{\ ,}
\end{equation}
where $n_{\rm e}$ is the number density of free electrons, $m_{\rm
  grain}$ is the mass of the dust grain, $c_{\rm s}$ is the sound
speed in the gas, and $T$ is the ISM temperature.  The functions appearing in equation (1) are defined as follows:

\begin{equation}
G_{\rm 0}(s)  = \frac{8 s}{3 \pi^{\frac{1}{2}}} \left(1 + \frac{9
    \pi}{64} s^2 \right)^{\frac{1}{2}} \mbox{\ ,}
\end{equation}

\begin{equation}
G_{\rm 2}(s)  = s \left( \frac{3}{4} \pi^{\frac{1}{2}} + s^3
\right)^{-1}  \mbox{\ .}
\end{equation}

Finally, for the electrostatic potential $U$ appearing in equation
(2), which depends on the grain charge and dictates the magnitude of the Coulomb force, we adopt
a wide range of values corresponding to various grain charges.  While
the grain charge is in part set by the photoionization rate of grains
due to radiation emitted from the star undergoing an accretion event,
it is also a function of the interstellar radiation field, which can
be either more or less intense than the stellar radiation field,
depending on the environment in which an accretion event occur.  It also
depends strongly on the number density of free electrons and ions in
the ISM, which accrete onto grains and alter their charge (see e.g. 
Weingartner \& Draine 2001c; Akimkin et al. 2015).  Given that the
radiation field and ionization state of the ISM during accretion events, which may have occurred at any point in the
$\sim$ 13 Gyr history of a typical low-mass Pop~III star, are unknown and
very uncertain, we consider a range of possible grain
charges and ISM ionization states in evaluating $F_{\rm drag}$, as shown in Figure 5.
This Figure shows the value of the quantity $G_{\rm 3}$ defined below
and appearing in the right side of equation (1):

\begin{equation}
G_{\rm 3} =  G_{\rm 0}(s) + f_{\rm e} \psi^2
  {\rm ln}(\Lambda) G_{\rm 2}(s)  \mbox{\ .}
\end{equation} 

We show the value of $G_{\rm 3}$ for a large
range of these values in Figure 5.  For this plot, we have assumed an
ISM temperature of $T$ = 50 K and a free electron number density of
$n_{\rm e}$ = 1 cm$^{-3}$, with the value of $G_{\rm 3}$ depending only relatively weakly on these
values.  Large values of $G_{\rm 3}$, corresponding to large grain charge
and/or free electron fraction, reflect a strong drag force (equation
1).  This, in turn, can dramatically decrease the likelihood of dust
segregation, as we discuss next.

  \begin{figure}
   \includegraphics[angle=0,width=3.4in]{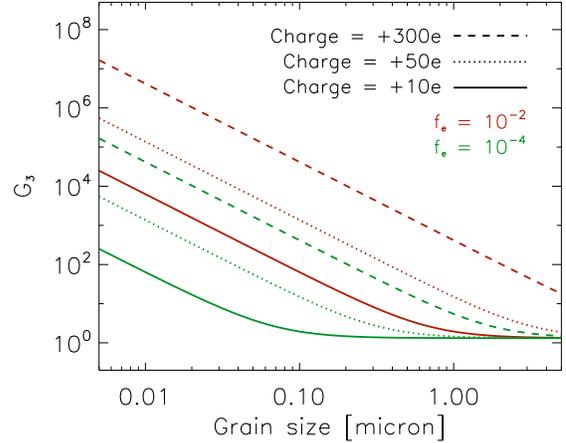}
   \caption{The factor $G_{\rm 3}$ appearing in equation (1) which
     encapsulates the impact of collisional gas drag and Coulomb drag
     on the  the critical densities $n_{\rm
       H,crit}$ shown in Figures 6, 7 and 8.  For values of $G_{\rm
       3}$ close to unity, the force due to collisional gas drag
     dominates that due to Coulomb drag.  Given the large
     uncertainties associated with the environments in which accretion
   may occur onto low-mass Pop~III stars, here we show $G_{\rm 3}$ for
 a variety of grain charges and ISM free electron fractions, as
 labelled.  For large grain charge and/or large free electron fraction
 $f_{\rm e}$, likely associated with intense
 radiation fields, its value is large, implying low
 values of the critical densities shown in Figures 6, 7 and 8.}
   \end{figure}

\section{The critical density for dust segregation}
From the definition we have adopted for the Eddington ratio $f_{\rm Edd}$ for dust, the net outward force $F_{\rm rad}$ on a dust grain due to
radiation from the star is given by 

\begin{equation}
F_{\rm rad} = (f_{\rm Edd} - 1) \frac{G M_{\rm *} m_{\rm grain}}{r^2} \mbox{\ ,}  
\end{equation}
where $m_{\rm grain}$ is
the mass of the grain, $M_{\rm *}$ is the mass of the star,
and $r$ is the distance from the star; the last term on the right
side is simply the definition of the gravitational force on the grain.
If the magnitude of the 
radiation force (equation 8) exceeds that of the drag force (equation
1), then a dust outflow will be set up
at the Bondi radius of the star.  If this condition is satisfied, then the 
dust grains will not be accreted, even while the gas is
freely accreting onto the star.  In particular, we 
define the critical density of hydrogen atoms $n_{\rm H}$ in equation
(1) as that above which the gas drag force exceeds the outward radiative
force, in which case the dust grains are entrained in the accretion flow.

Figures 6, 7 and 8 show the grain size-dependent values we find for
this critical density, for the Pop~III MS, Pop~III RG and
Pop~II protostellar cases, respectively.  In each Figure, a fiducial ISM
sound speed of $c_{\rm s}$ = 0.7 km s$^{-1}$ is assumed.  The Eddington ratios
shown in Figures 2, 3 and 4 are also adopted.  

\begin{figure}
   \includegraphics[angle=0,width=3.4in]{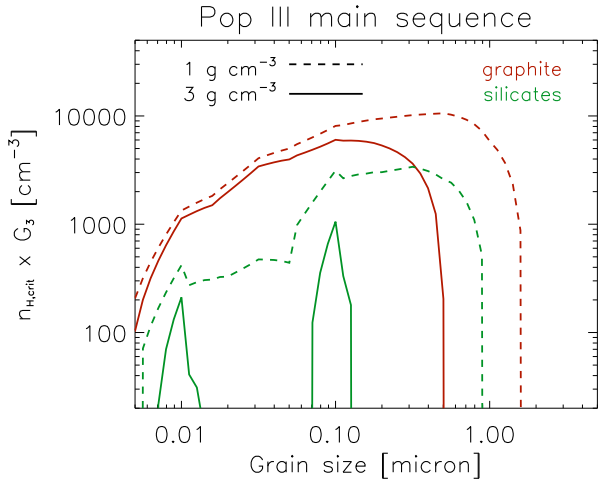}
   \caption{The critical ISM density above which gas drag entrains dust in an
     accretion flow, for the case of Bondi-Hoyle accretion with an ISM
     sound speed of $c_{\rm s}$ = 0.7 km s$^{-1}$, shown as a
     function of grain size (following from equations 1 and 7), for the case of a
     Pop~III MS star.  For a given grain size, less dense grains ({\it dashed lines}) experience a smaller gravitational force than more dense grains ({\it solid lines}), resulting in larger Eddington ratios and larger critical densities.
      The dependence of the critical density on the ISM ionized
fraction and grain charge, which dictate the
strength of the Coulomb drag force, is shown in Figure 5.}
   \end{figure}

For cases in which the Eddington ratio is less than unity, there is no
well-defined critical density, since the grains are accreted
regardless of the effect of gas drag.  This is the case, in
particular, for compact (high density) grains in the Pop~III MS case 
shown in Figure 5.  Following the trends described in Section 2
for the Eddington ratios, less dense grains are less susceptible to
gas drag entraining them in accretion flows, due to their higher
surface area-to-mass ratios.  

  \begin{figure}
   \includegraphics[angle=0,width=3.4in]{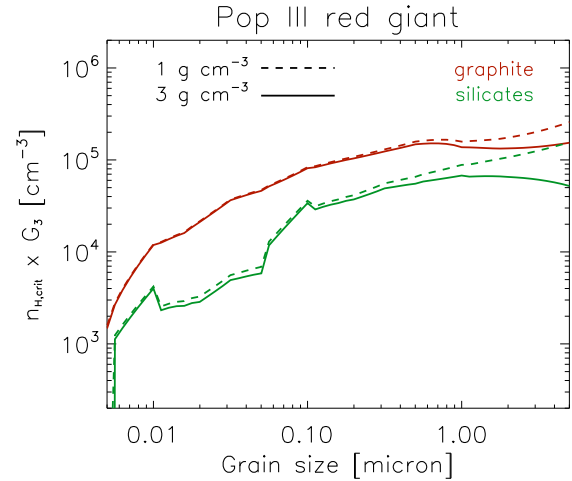}
   \caption{Just as Figure 6, but for the case of a Pop~III RG.}
   \end{figure}

  \begin{figure}
   \includegraphics[angle=0,width=3.4in]{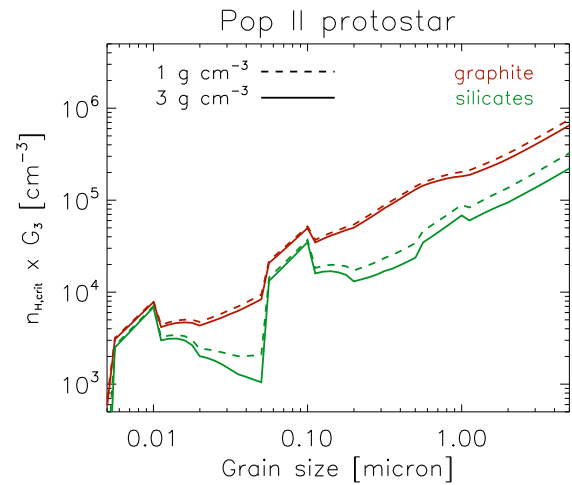}
   \caption{Just as Figures 6 and 7, but for the case of a Pop~II protostar.}
   \end{figure}

For cases of relatively low grain charge and/or ISM free electron fraction,
the critical densities shown in the Figures are, in general, much
larger than the densities of the gas clouds filling the vast majority
of the Galaxy (see e.g. Talbot \& Newman 1977; Rosolowsky 2010).  
Therefore, it is likely that dust segregation would occur during accretion onto low-mass Pop~III stars in
regions with relatively weak radiation fields and/or
weak ionization.  However, the elevated grain-ionizing photon flux in
the vicinity of massive stars can lead to large grain charges of up to 
$\sim$ 200 e+, as shown recently by Akimkin et al. (2015; see also
Gail \& Sedlmayr 1979);  such a
large grain charge could lead to critical densities as low as $n_{\rm
  H,crit}$ $<$ 1 cm$^{-3}$ for relatively small grains, depending on
the ionization state of the ISM.  If subjected to a weaker
interstellar radiation field (e.g. Weingartner \&
Draine 2001c)
the grain charge is likely to be much lower, leading to smaller
values of $G_{\rm 3}$ $\la$ 10$^2$ and critical densities $n_{\rm H,crit}$ $\ga$
10$^2$ cm$^{-3}$.  If the grains are charged predominantly due to
photoionization by accreting low-mass Pop~III stars themselves, the
grain charge is likely to be intermediate between these two cases, the
exact value depending on grain size and composition, as well as on
the ISM density, temperature and ionization state.

For simplicity, in the next Section we shall estimate the
chemical signatures of stars polluted via accretion under the
assumption that the ISM density is lower than the critical values
shown in the Figures.  This will suffice to outline the qualitative
trends that we expect dust segregation to imprint on the chemical
signatures of accretion-polluted low-mass stars.  We emphasize,
however, that especially in regions with intense radiation fields and/or strong
ISM ionization, the critical density may be lower than the ISM
density, making dust segregation unlikely to occur.

\begin{table*}
\begin{center}
\begin{tabular}{ccccccccccc}
 \hline \\ [-1.5ex]
            &  [C/Fe] & [N/Fe] & [O/Fe] & [Mg/Fe] & [Si/Fe] & [Ti/Fe] & [Cr/Fe] & [Mn/Fe] & [Ni/Fe] & [Zn/Fe] \\
 \\ [-1.5ex] \hline \\ [-1.5ex]
$\rho_{\rm grain}$ = 3 g cm$^{-3}$ \\
\\ [-1.5ex] \hline \\ [-1.5ex]
Pop III MS     & -0.03 & 0.15 & 0.12   & 0.01 & 0.009 & -0.001 & -0.0001 & 0.003 & -.0.0005 & 0.06 \\ 
Pop III RG    & 2.02 & 2.13 & 2.0  & 0.97 & 0.87  & -0.89 & -0.04 & 0.47 & -0.19 & 1.69 \\  
Pop II protostar & 2.02 & 2.13 & 2.0 & 0.97 & 0.87 & -0.89 & -0.04 & 0.47 & -0.19 & 1.69 \\
\\ [-1.5ex] \hline \\ [-1.5ex]
$\rho_{\rm grain}$ = 1 g cm$^{-3}$ \\
 \\ [-1.5ex] \hline \\ [-1.5ex]
Pop III MS & 2.02 & 2.13 & 2.0  & 0.97 & 0.87  & -0.89 & -0.04 & 0.47 & -0.19 & 1.69 \\  
Pop III RG & 2.02 & 2.13 & 2.0  & 0.97 & 0.87  & -0.89 & -0.04 & 0.47 & -0.19 & 1.69 \\ 
Pop II protostar & 2.02 & 2.13 & 2.0 & 0.97 & 0.87 & -0.89 & -0.04 & 0.47 & -0.19 & 1.69 \\
\\ [-1.5ex] \hline \\ [-1.5ex]
\end{tabular}
\caption{The expected chemical signatures of 0.8 M$_{\odot}$ Pop III
  MS and RG stars, and of 0.8 M$_{\odot}$ Pop~II protostars, due to accretion of material from the ISM having a solar
  abundance pattern and dust depletion properties as observed in the
  solar neighborhood.
  The abundance ratios are given for dust densities of $\rho_{\rm
    grain}$ = 1 and 3 g cm$^{-3}$, in the bottom and top rows, respectively.  The same distinct chemical signature, directly imprinted from the dust depletion properties of the local ISM, is predicted in nearly all cases.  It is only the Pop~III MS star in the case of compact (high density) grains, for which the Eddington ratio drops below unity (see Figure 2), that does not disply this signature.}
\end{center}
\end{table*}

\section{The Expected Chemical Signature}
To derive the expected chemical signatures of stars polluted with
metals via accretion of gas and/or dust from the ISM, we must take
into account the size distribution of dust grains and the depletion
factors of the various elements onto dust grains.
Here we consider grain sizes (for both graphite and silicates) defined by the
commonly used, power law size distribution presented in Mathis,
Rumpl \& Nordsieck (1977; MRN), in which the number of grains $N$ with
a given size $r_{\rm grain}$ is given by $dN/dr_{\rm grain}$ $\propto$
$r^{-3.5}_{\rm grain}$ and the minimum and maximum grain sizes are 5
nm and 250 nm, respectively.\footnote{This is a simple choice, which is suitable owing to the
large uncertainties elsewhere in our calculations, which include the
uncertainties in the dust fraction, composition and density of the
accreting ISM.  More refined grain size distributions for the local
ISM include those presented by Draine \& Lee
(1984) and Weingartner \& Draine (2001a).}
We use the element-dependent depletion factors derived for the
local ISM, as presented in Jenkins (2009).\footnote{We have adopted
  values that have been inferred for the cold, neutral ISM.  Lower
  depletion factors have also been inferred for the more diffuse ISM
  (see e.g. Jenkins 2009).}  For simplicity, and
consistent with previous work (e.g. Field 1974; Draine 2003a; Chiaki et
al. 2014), we assume all depleted carbon to be in graphite and silicates to comprise all 
other depleted elements.

To obtain the chemical signatures, we sum up all of the elements that
would be accreted in the form of gas and those that reside in dust
grains which have Eddington ratios below unity.  We do not include the
elements residing in dust grains with Eddington ratios greater than
unity, as these are assumed to not be accreted.  Finally, for
simplicity we assume that the chemical signature is due solely to the
accretion of material from an ISM with solar abundance ratios of heavy
elements (i.e. those heavier than hydrogen and helium) and dust
depletion properties as observed in the solar neighborhood.  While
this allows for concrete predictions, since these are observed
quantities, it is also likely that accretion onto low-mass Pop~III stars
and onto the earliest Pop~II protostars could have occurred from an
ISM with somewhat different heavy element abundance ratios and dust
depletion properties (see e.g. Chiaki et al. 2014; Ritter et al. 2014).

We present the expected chemical signatures in Table 1, 
for dust grain densities of 1 and 3 g cm$^{-3}$. 
For each grain density, we present the results that we find for
the cases of 0.8 M$_{\odot}$ MS Pop III
stars, Pop III RGs, and Pop II protostars.  For the Pop III cases,
accretion is assumed to occur from the ISM, while for the Pop II
protostar case it is assumed to take place during the growth of the
protostar itself via accretion from a molecular cloud.
We consider the abundances of ten elements relative to iron, as
labeled in the Table.

For cases in which the Eddington ratios of all dust grains are below
unity (as shown in Figures 2, 3 and 4), all dust and gas are accreted 
onto the star and the abundance ratios are solar (i.e. for an element
X, [X/Fe] = 0).  This is nearly the case for the
Pop~III MS star we consider, for dense grains (i.e. $\rho_{\rm grain}$
$\simeq$ 3 g cm$^{-3}$), and hence the abundance ratios are all near solar values.  
The more luminous Pop~III RGs and Pop~II
protostars, however, tend to have higher Eddington ratios, which
result in a distinctive abundance pattern.  In these cases, where the
Eddington ratio is greater than unity for all grains, no dust
is accreted onto the star and the abundance ratios follow directly
from the dust depletion properties.  This distinctive abundance
pattern emerges for the cases of Pop~II protostars and Pop~III
RGs, as well as for Pop~III MS stars given sufficiently low grain densities.  
We note that this distinct signature predicted for Pop~II
protostars would likely be complicated by the fact that some amount
of dust will have gone into the protostar at its initial collapse,
before it begins radiating appreciably.  Also, as noted in the last
Section, such protostars may accrete from molecular clouds that may
have densities higher than the critical densities shown in Figure 8.

We emphasize that the abundance ratios listed in Table 1 are for the specific case
of an ISM with heavy element abundance ratios similar to those
observed in the Sun and for dust depletion properties as observed in
the local ISM.  In reality, it is very likely that accretion onto
low-mass Pop~III stars could occur from an ISM that is different than
this.  Likewise, the earliest Pop~II protostars likely accrete gas
from an ISM that is different than this.  That said,
it is a robust conclusion that due to dust segregation in accretion
flows elements that are heavily depleted onto dust grains are not as
readily accreted as elements which are in the gas phase.  In turn,
this implies a robust chemical signature with elements largely
in the gas phase being more abundant than elements that are largely
depleted. 

As mentioned above, the most clear signatures of dust-segregated accretion are present in cases in
which the Eddington ratio exceeds unity for all grains, and no dust is
accreted onto the star.  
In this case, since iron is almost completely depleted onto dust grains in
the local ISM and carbon and oxygen are much less depleted (see
e.g. Table 4 of Jenkins 2009), this implies large characteristic
values of [C/Fe] = 2.02, [N/Fe] = 2.13, and [O/Fe] = 2.0.  In
turn, as magnesium and silicon are more depleted than carbon and
oxygen, but still less depleted than iron, characteristic values for these
elements are somewhat lower:   [Mg/Fe] = 0.97 and [Si/Fe] = 0.87.
Also, due to the high depletion factor of titanium, its expected
abundance ratio is quite low, [Ti/Fe] = -0.89.  Finally, the low depletion of 
zinc leads to a high expected abundance ratio of [Zn/Fe] = 1.69.

Many of these predicted abundance ratios are, in fact, similar
to the abundance ratios that have been inferred for a large range of
metal-poor stars.  In particular, the carbon-enhanced metal-poor
(CEMP) stars represent a large fraction of the most metal-poor stars
known (e.g. Aoki et al. 2013; Lee et al. 2013; Yong
et al. 2013; Carollo et al. 2014; Placco et al. 2014, 2015; Schlaufman \&
Casey 2014; Bonafacio et al. 2015; Frebel et al. 2015), and in many cases exhibit large carbon abundances relative to
iron such as those shown in Table 1.  There have been
reported stars in the literature with abundances that are similar to
those we find for other elements shown, as well.  

In Figure 9, we compare our
predicted chemical signature (for the case of no dust accretion) with
the data on a selection of CEMP stars from the literature (Aoki et al. 2007; Norris et al. 
2013; Spite et al. 2013).     
As the Figure shows, for many elements the data qualitatively match
the expected chemical signature, although there are outliers.  In
particular, the predicted low value for [Ti/Fe] is not found in the
data, nor is the predicted high value for [Zn/Fe].  Nonetheless, 
based on the broad agreement with the data that we find over the full range of elements, 
we tentatively conclude 
that some fraction of observed CEMP stars could, in fact, be Pop~III
stars that have been polluted due to accretion of material from the ISM of the Galaxy. 
We emphasize that any star that matches well all the abundance ratios
shown in Figure 9, including
those for titanium and zinc, would be a very strong candidate
polluted Pop~III star.  That said, given the large uncertainties in
the ISM and dust properties over the history and spatial extent
(e.g. Ochsendorf \& Tielens 2015) of the Galaxy and its
lower-metallicity progenitors (see e.g. Tchernyshyov et al. 2015), it is
also difficult to rule out this possibility for any of the stars
shown in Figure 9 solely on the basis of the quality of match to this specific predicted 
chemical signature.

  \begin{figure}
   \includegraphics[angle=0,width=3.4in, trim=40 0 0 0,clip]{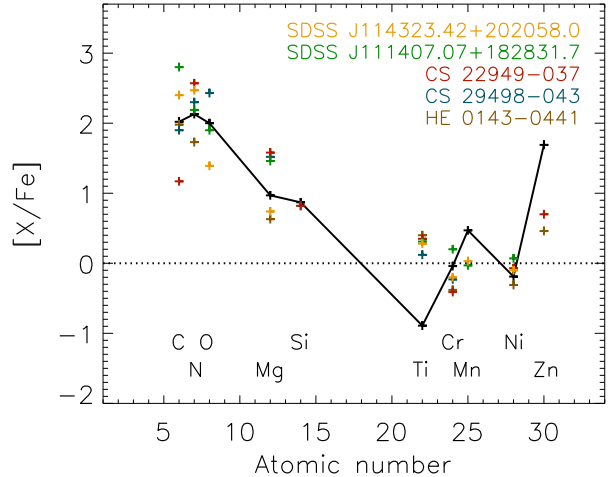}
   \caption{The expected chemical signature of accretion from the
     local ISM, for the case of no dust accretion, for the
     same ten elements shown in Table 1 ({\it black points and
       lines}), and the data for these elemental abundances for five 
       CEMP stars reported in the literature, each color-coded separately as labeled.  
       While the data cluster
       around the expected chemical abundances for most elements, in particular those with low atomic numbers,
       there are heavier elements which are outliers.  The weak and strong
       depletion of zinc and titanium, respectively, lead to
       distinctive signatures which are not particularly well fit by the data.}
   \end{figure}

\section{Conclusions and Discussion}
We have estimated the impact of the dynamics of dust grains in
accretion flows on the chemical abundances of low-mass 
Pop~III stars and Pop~II protostars, and we have
outlined the conditions in which distinct chemical signatures could be
left due to the segregation of dust from the accreting gas.  Based on
these results, we have predicted the expected abundance patterns for low-mass Pop~III
stars and Pop~II protostars that accrete gas from an ISM with
heavy element abundance ratios similar to those of the Sun and with
dust depletion properties similar to those observed in the local ISM. 
Our main conclusions are as follows:

\begin{itemize}

\item Due to the pressure of the radiation emitted from low-mass stars, 
  dust grains can be segregated from the gas in accretion flows
  and can therefore be prevented from accreting onto such objects.

\item Dust segregation is more likely to occur for grains with lower
  densities (more porous grains) and in accretion flows onto
  relatively luminous low-mass stars, such as Pop~III red giants and
  Pop~II protostars in particular.  

\item Dust segregation is less likely to occur in dense accretion
  flows, however, due to the larger rate of collisions with accreting
  gas particles.  It is also less likely to occur in strongly ionized regions of the
  ISM or in regions subject to intense radiation fields, due to the
  Coulomb drag on charged dust grains.

\item As the distinct abundance ratios that we predict to result from
  dust-segregated accretion from the ISM are in
  broad agreement with those found for many observed CEMP stars, it appears
  possible that some fraction of observed CEMP stars may be
  Pop~III stars that have been polluted by accretion from the ISM of
  the Galaxy.

\item Two distinct chemical signatures that may particularly
  strongly indicate a Pop~III origin for CEMP stars are a low
  titanium abundance relative to iron, and a large zinc abundance
  relative to iron, as shown in Figure 9.

\end{itemize}

We emphasize that there are other possible explanations for the
origin of CEMP stars.  Among these are fall-back in the first
supernovae resulting in the preferential ejection of carbon over iron
(e.g. Iwamoto 2005; see also Sluder et al. 2015), mass transfer from companion stars (e.g. Lee
et al. 2014), atomic line cooling by carbon and oxygen as the critical
process leading to the first low-mass stars (Bromm \& Loeb 2003), iron
depletion in the formation of a circumstellar disk (e.g. Venn et
al. 2014), stellar winds from fast-rotating massive stars (Maeder et
al. 2014), and
the preferential ejection of iron from the low-mass dark matter haloes in which
the first stars formed (Cooke \& Madau 2014).  Conversely, we also note
that low-mass Pop~III stars may not exhibit the chemical signatures of
accretion from the ISM at all, if they are able to repel interstellar material by
launching solar-like winds (Johnson \& Khochfar 2011), as observations
suggest many low-mass stars may do (e.g. Torres et al. 2015).  In this case,
the primordial nature of such stars would be readily apparent.

\section*{Acknowledgements}
Work at LANL was done under the auspices of the National Nuclear Security 
Administration of the US Department of Energy at Los Alamos National 
Laboratory under Contract No. DE-AC52-06NA25396.  JLJ would like to
thank George Becker, Joe Smidt and Joyce Guzik for helpful discussion, as well as Thomas Greif and Volker
Bromm for valuable feedback on an early version of this work.  The author
is thankful to an anonymous reviewer for helpful comments on the role of Coulomb
drag in coupling gas and dust grains, as well as to another anonymous
reviewer for insightful comments and suggestions.


\appendix

\section{The possibility of dust sublimation}
It is possible that dust grains could be sublimated in the accretion
processes that we consider, in which case their constituent elements
would enter the gas phase and could be accreted.
In particular, sublimation is expected to occur for most grains at
temperatures $\ga$ 10$^3$ K (e.g. Kobayashi et al. 2011).  
We can estimate the temperature of dust grains in an accretion flow, 
by assuming that they are in thermal equilibrium, i.e., that the grains
radiate thermal energy at the same rate that they absorb radiative energy from
the star.  Taking it that all radiation incident on the grain is
absorbed, which provides an upper limit to the heating rate of the grain, 
we find that the equilibrium dust temperature is 

\begin{equation}
T_{\rm dust} \simeq 3 \times 10^2 \, {\rm K} \left(\frac{L_{\rm *}}{L_{\odot}}\right)^{\frac{1}{4}} \left(\frac{r}{1 \, {\rm AU}}\right)^{-\frac{1}{2}}  \mbox{\ ,}
\end{equation} 
where $L_{\rm *}$ is the luminosity of the star and $r$ is the
distance from the star.  While the low-mass
Pop~III stars and Pop~II protostars 
of interest here may have luminosities up to $\sim$ 100 times larger than that
of the Sun, dust is likely to be sublimated only once it comes within $\sim$ 1 AU of
the star.  As this is orders of magnitude closer than the Bondi radius $r_{\rm Bondi}$
of the star, within which the dust would be segregated from the gas if the 
Eddington ratio for dust exceeds unity, it appears unlikely that dust grains 
that would not accrete onto the star could be destroyed due to sublimation.


\begin{thebibliography}{99}
\bibitem[2(2000)]{b}Akimkin, V.~V., Kirsanova, M.~S., Pavlyuchenkov,
  Y.~N., Wiebe, D.~S. 2015, MNRAS, 449, 440
\bibitem[2(2000)]{b}Aoki, W., et al. 2007, ApJ, 655, 492
\bibitem[2(2000)]{b}Aoki, W., et al. 2013, AJ, 145, 13
\bibitem[2(2000)]{b}Baines, M.~J., Williamsn, I.~P., Asebiomo,
  A.~S. 1965, MNRAS, 130, 63
\bibitem[2(2000)]{b}Beers, T.~C., Christlieb, N. 2005, ARA\&A, 43, 531
\bibitem[2(2000)]{b}Bonafacio, P., et al. 2015, A\&A, in press (arXiv:1504.05963)
\bibitem[2(2000)]{b}Bond, H.~E. 1981, ApJ, 248, 606
\bibitem[2(2000)]{b}Bromm, V. 2013, RPPh, 76, 2901
\bibitem[2(2000)]{b}Bromm, V., Loeb, A. 2003, Nat, 425, 812
\bibitem[2(2000)]{b}Caffau, E., et al. 2011, Nat, 477, 67
\bibitem[2(2000)]{b}Carollo, D., et al. 2014, ApJ, 788, 180
\bibitem[2(2000)]{b}Cayrel, R., et al. 2004, A\&A, 416, 1117
\bibitem[2(2000)]{b}Chiaki, G., et al. 2014, MNRAS, submitted (arXiv:1410.8384)
\bibitem[2(2000)]{b}Clark, P.~C., Glover, S.~C.~O., Klessen, R.~S.,
  Bromm, V. 2011, ApJ, 727, 110
\bibitem[2(2000)]{b}Cooke, R.~J., Madau, P. 2014, ApJ, submitted (arXiv:1405.7369)
\bibitem[2(2000)]{b}Dopcke, G., Glover, S.~C.~O., Clark, P.~C., Klessen, R.~S. 2013, ApJ, 766, 103
\bibitem[2(2000)]{b}Draine, B.~T., Lee, M.~L. 1984, ApJ, 285, 89
\bibitem[2(2000)]{b}Draine, B.~T. 2003a, ARA\&A, 41, 241
\bibitem[2(2000)]{b}Draine, B.~T. 2003b, arXiv:0304488
\bibitem[2(2000)]{b}Draine, B.~T., Salpeter, E.~E. 1979, ApJ, 231, 77 
\bibitem[2(2000)]{b}Draine, B.~T. 2004, 'Astrophysics of Dust in Cold
  Clouds', arXiv:0304488
\bibitem[2(2000)]{b}Dullemond, C.~P., Dominik, C. 2005, A\&A, 434, 971
\bibitem[2(2000)]{b}Dunham, M.~M., et al. 2014, PRPL, accepted (arXiv:1401.1809)
\bibitem[2(2000)]{b}Dwek, E. 1998, ApJ, 501, 643
\bibitem[2(2000)]{b}Field, G.~B. 1974, ApJ, 187, 453
\bibitem[2(2000)]{b}Frebel, A. 2010, AN, 331, 474
\bibitem[2(2000)]{b}Frebel, A., Simon, J.~D., Kirby, E.~N. 2014, ApJ,
  786, 74
\bibitem[2(2000)]{b}Frebel, A., Chiti, A., Ji, A.~P., Jacobson, H.~R.,
  Placco, V.~M. 2015, ApJ, submitted (arXiv:1507.01973)
\bibitem[2(2000)]{b}Frebel, A., Johnson, J.~L., Bromm, V. 2009, MNRAS,
  392, L50
\bibitem[2(2000)]{b}Gail, H.-P., Sedlmayr, E. 1979, A\&A, 77, 165
\bibitem[2(2000)]{b}Gao, L., Theuns, T., Frenk, C.~S., Jenkins, A., Helly, J.~C., Navarro, J., Springel, V., White, S.~D.~M. 2010, MNRAS, 403, 1283
\bibitem[2(2000)]{b}Greif, T., et al. 2011, ApJ, 737, 75
\bibitem[2(2000)]{b}Greif, T. 2014, arXiv:1410.3482
\bibitem[2(2000)]{b}Hartwig, T., Bromm, V., Klessen, R.~S., Glover, S.~C.~O. 2014, MNRAS, submitted (arXiv:1411.1238)
\bibitem[2(2000)]{b}Heng, K., Draine, B.~T. 2009, arXiv:0906.0773
\bibitem[2(2000)]{b}Henry, R.~J.~W. 1970, ApJ, 161, 1153
\bibitem[2(2000)]{b}Hirano, S., et al. 2014, ApJ 781, 60
\bibitem[2(2000)]{b}Iat{\' i}, M.~A., et al. 2001, MNRAS, 322, 749
\bibitem[2(2000)]{b}Iben, I. 1983, MmSAI, 54, 321
\bibitem[2(2000)]{b}Ivezic, Z., Elitzur, M. 1995, ApJ, 445, 415
\bibitem[2(2000)]{b}Ivezic, Z., Elitzur, M. 2010, MNRAS, 404, 1415
\bibitem[2(2000)]{b}Iwamoto, N., et al. 2005, Sci, 309, 4511
\bibitem[2(2000)]{b}Jenkins, E.~B. 2009, ApJ, 700, 1299
\bibitem[2(2000)]{b}Johnson, J.~L., Khochfar, S. 2011, MNRAS, 413, 1184
\bibitem[2(2000)]{b}Karlsson, T., Bromm, V., Bland-Hawthorn, J. 2013, ReMP, 85, 809
\bibitem[2(2000)]{b}Keller, S.~C., et al. 2014, Nat, 506, 463
\bibitem[2(2000)]{b}Kirby, E.~N., et al. 2011, ApJ, 727, 78
\bibitem[2(2000)]{b}Kobayashi, H., et al. 2011, Earth Planet. Space,
  accepted (arXiv:1104.5627)
\bibitem[2(2000)]{b}Komiya, Y., Habe, A., Suda, T., Fujimoto,
  M.~Y. 2010, ApJ, 717, 542
\bibitem[2(2000)]{b}Komiya, Y., Suda, T., Fujimoto, M.~Y. 2015, ApJ,
  submitted (arXiv:1507.01664)
\bibitem[2(2000)]{b}Laor, A., Draine, B.~T. 1993, ApJ, 402, 441L
\bibitem[2(2000)]{b}Lee, Y.~S., et al. 2013, AJ, 146, 132
\bibitem[2(2000)]{b}Lee, Y.~S., et al. 2014, ApJ, 788, 131
\bibitem[2(2000)]{b}Machida, M.~N., Nakamura, T. 2015, MNRAS, 448, 1405
\bibitem[2(2000)]{b}Madau, P., Kuhlen, M., Diemand, J., Moore, B., Zemp, M., Potter, D., Stadel, J. 2008, ApJ, 689, L41
\bibitem[2(2000)]{b}Maeder, A., Meynet, G., Chiappini, C. 2014, A\&A,
  in press (arXiv:1412.5754)
\bibitem[2(2000)]{b}Marigo, P., Girardi, L., Chiosi, C., Wood,
  P.~R. 2001, A\&A, 371, 152
\bibitem[2(2000)]{b}Mathis, J.~S., Rumpl, W., Nordsieck, K.~H. 1977,
  ApJ, 217, 425
\bibitem[2(2000)]{b}Mathis, J.~S. 1996, ApJ, 472, 643
\bibitem[2(2000)]{b}Mukai, T., Ishimoto, H., Kozasa, T., Blum, J.,
  Greenberg, J.~M. 1992, A\&A, 262, 315
\bibitem[2(2000)]{b}Netzer, N. 2007, BaltA, 16, 120
\bibitem[2(2000)]{b}Norris, J.~E. 2013, ApJ, 762, 28
\bibitem[2(2000)]{b}Ochsendorf, B.~B., Tielens, A.~G.~G.~M. 2015,
  A\&A, submitted (arXiv:1501.02256)
\bibitem[2(2000)]{b}Ossenkopf, V. 1993, A\&A, 280, 617
\bibitem[2(2000)]{b}Picardi, I., Chieffi, A., Limongi, M., Pisanti,
  O., Miele, G., Mangano, G., Imbriani, G. 2004, ApJ, 609, 1035
\bibitem[2(2000)]{b}Placco, V.~M., Frebel, A., Beers, T.~C.,
  Stancliffe, R.~J. 2014, ApJ, submitted (arXiv:1410.2223)
\bibitem[2(2000)]{b}Placco, V.~M., et al. 2015, ApJ, accepted (arXiv:1507.03656)
\bibitem[2(2000)]{b}Ritter, J.~S., Sluder, A., Safranek-Shrader, C.,
  Milosavljevi{\' c}, M., Bromm, V. 2014, MNRAS, submitted (arXiv:1408.0319)
\bibitem[2(2000)]{b}Rosolowsky, E., et al. 2010, ApJS, 188, 123
\bibitem[2(2000)]{b}Schaerer, D. 2002, A\&A, 382, 28
\bibitem[2(2000)]{b}Schlaufman, K.~C., Casey, A.~R. 2014, ApJ,
  accepted (arXiv:1409.4775)
\bibitem[2(2000)]{b}Shen, Y., Draine, B.~T., Johnson, E.~T. 2008, ApJ,
  689, 260
\bibitem[2(2000)]{b}Siess, L., Livio, M., Lattanzio, J. 2002, ApJ, 570, 329
\bibitem[2(2000)]{b}Simpson, I.~C., Simons, S., Williams, I.~P. 1980,
  Ap\&SS, 71, 3
\bibitem[2(2000)]{b}Sluder, A., Ritter, J.~S., Safranek-Shrader, C.,
  Milosavljevi{\' c}, M., Bromm, V. 2015, MNRAS, submitted (arXiv:1505.07126)
\bibitem[2(2000)]{b}Spite, M., Caffau, E., Bonifacio, P., Spite, F.,
  Ludwig, H.-G., Plez, B., Christlieb, N. 2013, A\&A, 552, 107
\bibitem[2(2000)]{b}Stacy, A., Bromm, V. 2014, ApJ, 785, 73
\bibitem[2(2000)]{b}Suda, T., Fujimoto, M.~Y., Itoh, N. 2007, ApJ, 667, 1206
\bibitem[2(2000)]{b}Susa, H., Hasegawa, K., Tominaga, N. 2014, ApJ,
  submitted (arXiv:1407.1374)
\bibitem[2(2000)]{b}Draine, B.~T., Sutin, B. 1987, ApJ, 320, 803
\bibitem[2(2000)]{b}Talbot, R.~J., Newman, M.~J. 1977, ApJS, 34, 295
\bibitem[2(2000)]{b}Tazaki, R., Nomura, H. 2015, ApJ, 799, 119
\bibitem[2(2000)]{b}Torres, G., et al. 2015, arXiv:1503.06310
\bibitem[2(2000)]{b}Tumlinson, J. 2010, ApJ, 708, 1398
\bibitem[2(2000)]{b}Tchernyshyov, K., et al. 2015, ApJ, submitted (arXiv:1503.08852)
\bibitem[2(2000)]{b}Venn, K.~A., et al. 2014, ApJ, 791, 98
\bibitem[2(2000)]{b}Voshchninnikov, N.~V., Il'in 1983, SvA, 27, 650
\bibitem[2(2000)]{b}Voshchninnikov, N.~V., Il'in, V.~B., Henning,
  T. 2005, A\&A, 429, 371
\bibitem[2(2000)]{b}Weingartner, J.~C., Draine, B.~T. 2001a, ApJ, 548, 296
\bibitem[2(2000)]{b}Weingartner, J.~C., Draine, B.~T. 2001b, ApJ, 553, 581
\bibitem[2(2000)]{b}Weingartner, J.~C., Draine, B.~T. 2001c, ApJS, 134, 263 
\bibitem[2(2000)]{b}Whitworth, A.~P., Bate, M.~R. 2002, MNRAS, 333, 679
\bibitem[2(2000)]{b}Yong, D., et al. 2013, ApJ, 762, 28
\bibitem[2(2000)]{b}Yoshii, Y. 1981, A\&A, 97, 280


\end{thebibliography}
\end{document}